\documentstyle[11pt,paspconf,epsf]{article}

\begin{document}

\title{UV Signposts of Resonant Dynamics in Disk Galaxies}

\author{William H. Waller\altaffilmark{1,2} and Christine M. Winslow}
\affil{Astronomy Program, Tufts University, Medford, MA 
02155}

\author{Michael N. Fanelli\altaffilmark{1}}
\affil{Department of Physics, University of North Texas,
Denton, TX 76203}

\author{Theodore P. Stecher}
\affil{NASA Goddard Space Flight Center, Code 681, Greenbelt, MD 20771}

\altaffiltext{1}{Senior Scientist, Raytheon STX Corporation} 
\altaffiltext{2}{Visiting Scientist, Harvard-Smithsonian Center for 
Astrophysics}



\begin{abstract}
Imaging in the restframe ultraviolet has proven to be an effective and
vital means of tracing dynamical patterns of star formation in 
galaxies out to high
redshifts.  Using images from the Ultraviolet Imaging Telescope (UIT),
Hubble Space Telescope (HST) and complementary groundbased 
telescopes, we have investigated the starburst activity and associated
dynamics in nearby early-type disk galaxies.  Concentrating on the 
starburst-ring (R)SA(r)ab galaxy M94 (NGC 4736), we find compelling
evidence for bar-mediated resonances as the primary drivers of
evolution at the present epoch.  Similar ring-bar dynamics may prevail
in the centers of early-type disk galaxies at high redshift.  The
gravitationally-lensed ``Pretzel Galaxy'' (0024+1654) at a redshift of
$\sim$1.5 provides an important precedent in this regard.
\end{abstract}




\section{UV Rings and Underlying Dynamics}
Inner rings and pseudo-rings are evident in 50\% of all spiral galaxies
and in more than 75\% of early-type barred spirals (Kormendy 1979; 
de Vaucouleurs \& Buta 1980; Buta \& Combes 1996).  These rings are often
characterized by intense starbirth activity.  UV imaging
neatly isolates such rings from the older disk and bulge components.  
UIT images of M100 (NGC 4321), NGC 1317, and M94 (NGC 4736) 
provide textbook examples of the starburst ring phenomenon (see  
http://trifle.gsfc.nasa.gov/UIT/Astro1/Astro1\_pictures.html).

\section{M94:  A Revealing Case Study}
As the closest early-type spiral galaxy of low inclination, M94 (NGC 4736)
has received concentrated attention from both observers and theorists.
This (R)SA(r)ab galaxy is noted for its inner ring of ongoing starburst
activity (R $\approx$ 1.0 kpc), oval stellar distribution at intermediate 
radius (R $\approx$ 4.9 kpc), and outer stellar ring near its Holmberg
radius (R $\approx$ 7.3 kpc).  

The UV images obtained by UIT and HST reveal hitherto unrecognized
patterns of recent star formation, whose presence lends further support to
the hypothesis of galaxy evolution via bar-mediated resonances.  The UIT's
FUV image shows the {\it starburst ring} 
in high contrast against a mostly dark
disk (Waller et al. 1997; 
see also http://trifle.gsfc.nasa.gov/UIT/Astro2/ngc4736\_pr.gif).  
Exterior to the ring are two 500-pc size
{\it bi-symmetric knots} on diametrically opposite sides of the nucleus.
The HST's NUV image of the central 20 arcsecs shows a 500-pc long {\it 
mini-bar} that had been previously inferred from photometric analyses of 
optical-band images.  By comparison, a groundbased R-band image shows the
underlying bulge and {\it oval disk} 
made up of cooler and typically older 
stars.

\begin{figure}
\plottwo{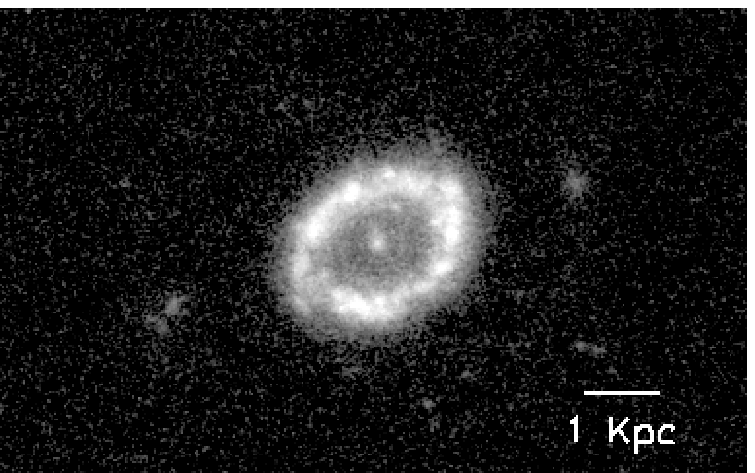}{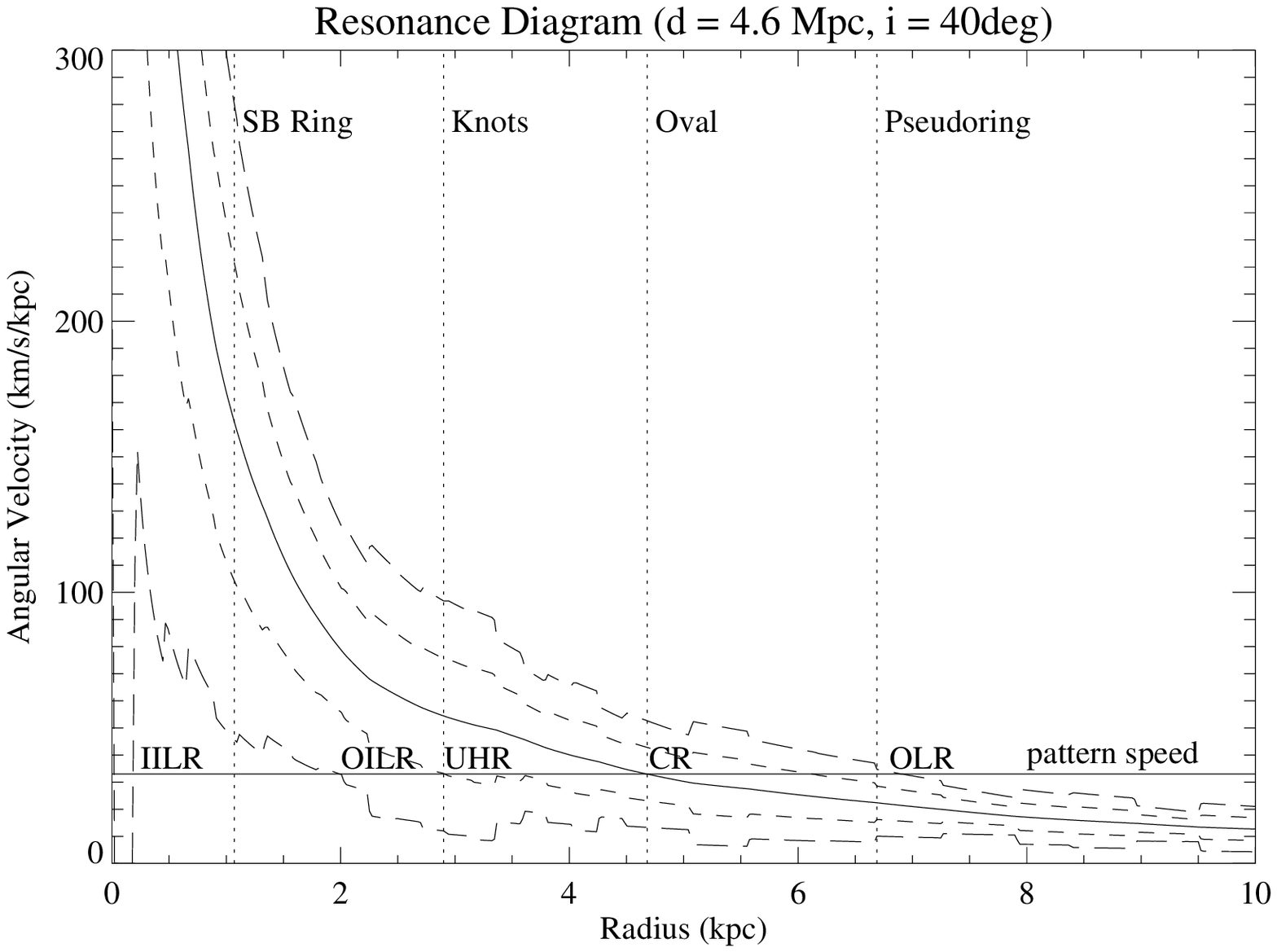}
\caption{Far-UV imaging of M94 by UIT reveals a starburst ring and 
two bi-symmetric knots of hot stars in high contrast against a mostly dark 
disk.  From the HI rotation curve of Mulder and van Driel (1993), the 
derived orbital resonance diagram for M94 shows key morphological features 
coincident with important resonances.  Here, the solid, short-dashed, and 
long-dashed lines respectively trace the angular frequencies $\Omega$, 
$\Omega \pm \kappa/4$, and $\Omega \pm \kappa/2$.  Radii corresponding to the
Inner-Inner Lindblad Resonance (IILR), Outer-Inner Lindblad Resonance (OILR),
Ultra-Harmonic Resonance (UHR),
Co-Rotation (CR), and Outer Lindblad Resonance (OLR) are noted.}\label{fig-1}
\end{figure}

Taken together, the nuclear minibar, starburst ring, bi-symmetric knots,
oval disk, and outer pseudo-ring provide compelling evidence for resonant
dynamics operating in the disk of M94.  More specifically, the radii of 
these various features match those of important orbital resonances, given
a pattern speed of 35 km s$^{-1}$ kpc$^{-1}$ and our adopted distance
and inclination (see Figure~\ref{fig-1}).  This pattern of orbital resonances
is most likely driven by some combination of the nuclear mini-bar and oval
distortion in the disk.  Note in Figure 1 that our fit places the
starburst ring between the two Inner Lindblad Resonances rather than 
aligned with the outer ILR as modeled by Gerin et al. (1991) and Mulder
\& Combes (1996).

\section{Resonant Rings at High Redshift?}
It is worth noting that at redshifts of 1--5, the 2-kpc diameter starburst 
ring in M94 would subtend angles of only 
(0.7$''$ -- 1.0$''$)H$_o$/75 in an 
Einstein-de Sitter Universe (q$_o$ = 1/2) and (0.3$''$ -- 0.2$''$)H$_o$/75
in an open (Milne) universe (q$_o$ = 0).  Therefore, many of the ``core-halo'' 
morphologies that are evident at high-redshift in the restframe FUV 
(cf. Giavalisco 1997) may, in 
fact be marginally-resolved representations of galaxies with starburst rings
in their centers.  Gravitationally-lensed galaxies are fortuitously
magnified, enabling resolution of their structure at high S/N.  
An important precedent in this regard is the gravitationally-lensed 
``Pretzel Galaxy'' which lies behind the galaxy cluster 0024+1654 at an
estimated redshift of 1.2 -- 1.8 (Colley et al. 1996; Tyson et al. 1997).
Detailed reconstructions of the multiply-lensed galaxy show a clear 
annular morphology on a scale of several kpc.  If M94 and other nearby 
ringed galaxies can be used as current-epoch analogues, the ``Pretzel Galaxy'' 
and perhaps other marginally-resolved ``core-halo'' galaxies at high redshift
may represent youthful inner disks evolving under the influence of oval
or bar asymmetries.  Conversely, if evidence for starburst rings at high
redshift proves to be sparse, then massive 
inner disks featuring ring-bar dynamics
have yet to form in most systems, or starbursting bulges are masking their
presence.

%
%

\acknowledgments

UIT research is funded through the Spacelab Office at NASA Headquarters 
under Project number 440-51.  W. H. Waller acknowledges partial support 
from NASA's Astrophysics Data Program (071-96adp). 

%
%

%

\end{document}